\documentclass[12pt]{article}
\renewcommand{\thefootnote}{\fnsymbol{footnote}}
\usepackage{epsfig}
\begin{document}
\newcommand{\be}{\begin{eqnarray}}
\newcommand{\dlq}{\lq\lq}
\newcommand{\ee}{\end{eqnarray}}
\newcommand{\ben}{\begin{eqnarray*}}
\newcommand{\een}{\end{eqnarray*}}
\newcommand{\stackeven}[2]{{{}_{\displaystyle{#1}}\atop\displaystyle{#2}}}
\newcommand{\lsim}{\stackeven{<}{\sim}}
\newcommand{\gsim}{\stackeven{>}{\sim}}
\newcommand{\un}[1]{\underline{#1}}
\renewcommand{\baselinestretch}{1.0}
\newcommand{\as}{\alpha_s}
\def\eq#1{{Eq.~(\ref{#1})}}
\def\fig#1{{Fig.~\ref{#1}}}
\begin{flushright}
NT@UW--02--021 \\
INT--PUB--02--41 \\
\end{flushright}
\vspace*{1cm} 
\setcounter{footnote}{1}
\begin{center}
{\Large\bf Correlation Functions and Cumulants~\\~~\\ 
in Elliptic Flow Analysis}
\\[1cm]
Yuri V.\ Kovchegov\footnote{e-mail: yuri@phys.washington.edu} $^{1}$ 
and  Kirill L.\ Tuchin\footnote{e-mail: tuchin@phys.washington.edu} $^{2}$ \\ ~~ \\ 
{\it $^1$ Department of Physics, University of Washington, Box 351560} \\ {\it
Seattle, WA 98195 } \\ ~~ \\ 
{\it $^2$ Institute for Nuclear
Theory, University of Washington, Box 351550 } \\ {\it Seattle, WA
98195 } \\ ~~ \\ ~~ \\
\end{center}
\begin{abstract}
We consider various methods of flow analysis in heavy ion collisions
and compare experimental data on corresponding observables to the
predictions of our saturation model proposed earlier \cite{KT}. We
demonstrate that, due to the nature of the standard flow analysis,
azimuthal distribution of particles with respect to reaction plane
determined from the second order harmonics should always be
proportional to $\cos 2(\phi-\Psi_R)$ independent of the physical
origin of particle correlations (flow or non-flow). The amplitude of
this distribution is always physical and proportional to
$v_2$. Two-particle correlations analysis is therefore a more reliable
way of extracting the shape of physical azimuthal anisotropy. We
demonstrate that two-particle correlation functions generated in our
minijet model of particle production \cite{KT} are in good agreement
with the data reported by PHENIX.  We discuss the role of non-flow
correlations in the cumulant flow analysis and demonstrate using a
simple example that if the flow is weak, higher order cumulants
analysis does not significantly reduce the contribution of non-flow
correlations to elliptic flow observable $v_2$ in RHIC data.
\end{abstract}
\renewcommand{\thefootnote}{\arabic{footnote}}
\setcounter{footnote}{0}

\newpage

\section{Introduction}

Differential elliptic flow in heavy ion collisions is defined as the
coefficient of the second harmonic of particle distribution with
respect to the reaction plane
\be\label{v2diff}
v_2 (p_T) \, = \, \left< \, \cos 2 (\phi_{p_T} - \Phi_R) \, \right>
\ee
where $\phi_{p_T}$ is the azimuthal angle of the produced particle
with transverse momentum $p_T$, $\Phi_R$ is the azimuthal angle of the
reaction plane and the brackets denote statistical averaging over all
particles with momentum $p_T$ and over different events. According to
emerging RHIC data differential elliptic flow is an increasing
function of $p_T$ for small $p_T$, which stops increasing and
saturates to a constant at $p_T \gsim 1.5 - 2$~GeV
\cite{starsat1,starsat2,starsat3}. While the hydrodynamic models are
successful in describing the low-$p_T$ increase of $v_2 (p_T)$
\cite{huov,teaney}, they disagree with the high-$p_T$ data. Models
incorporating jet quenching on top of hydrodynamic expansion also do
not agree with the high-$p_T$ data very well
\cite{gvw}. Describing the quark-gluon plasma (QGP) dynamics by 
covariant transport theory requires either extremely high initial
gluon density or very large parton-parton scattering cross sections in
order to fit the high-$p_T$ elliptic flow data \cite{denes}.

In our previous paper on the subject \cite{KT} we proposed a model of
non-flow particle correlations in the initial stages of heavy ion
collisions which described $v_2 (p_T)$ saturation data rather
well. The model was based on particle production mechanism in the high
energy regime when gluon and quark distribution functions of the
colliding nuclei reach saturation \cite{sat}. As was suggested by
McLerran and Venugopalan \cite{mv} the dominant gluon production
mechanism in the early stages of the collision is given by the
classical field of the nuclei. The field was found at the lowest order
in $\as$ in \cite{claa}. Extensive numerical simulations of the full
solution have been performed in \cite{kv} while an analytical ansatz
for the corresponding spectrum of the produced gluons was written down
in \cite{yuaa}. At the high energies achieved by RHIC experiments the
classical field alone can not account for particle production; thus
quantum corrections become important. For an ansatz of gluon
production cross section in AA including the nonlinear evolution of
\cite{bk} see \cite{KT1}.

The essence of our model proposed in \cite{KT} is the following: to
estimate the non-flow contribution to $v_2 (p_T)$ one has to calculate
the single and double inclusive gluon production cross sections first
in the framework of the simple McLerran-Venugopalan model and then
include the nonlinear evolution effects \cite{bk} in them. Two
produced gluons in the double inclusive cross section are of course
azimuthally correlated with each other \cite{KT}.  This correlation
can contribute to $v_2$ after being averaged over all particle pairs,
which is proportional to the total particle multiplicity squared. The
latter was related to the single inclusive gluon production cross
section similar to how it was done in
\cite{kln}. Two comments are in order here. First of all double gluon
production cross section of course can not be given by the classical
field and is therefore not a classical quantity. Calculating it thus
corresponds to the first (order $\as$) correction to
McLerran-Venugopalan model. Secondly the correlations in the double
inclusive gluon production cross section are not just back-to-back, as
one would naively expect. One of the main advantages of saturation
physics is that one does {\it not} have to assume that the momenta of
the produced gluons are large in order to use small coupling and twist
expansions like it is done in the collinear factorization
approach. Saturation calculations include all twists and the coupling
is kept small by the large saturation scale $Q_s$
\cite{mv,km,glmmt}. When the momenta of the produced two gluons are
not extremely large the correlations are not only back-to-back since
it is not required anymore by transverse momentum conservation as some
of the momentum can be carried away by other soft particles. A
calculation of the lowest order double inclusive cross section was
done in \cite{leostr} showing not only back-to-back ($\Delta \phi =
\pi$) but also collinear ($\Delta \phi = 0$) correlations.

Since the exact double inclusive gluon production cross section is not
known we constructed a simple model of single- and double- gluon
production \cite{KT} in the spirit of $k_T$-factorization approach
used in \cite{kln}. The model successfully described the saturation of
$v_2 (p_T)$ at high $p_T$ as well as centrality dependence of $v_2
(B)$. The goal of this paper is to discuss compatibility of our model
with other observables in the flow analysis.

There are other types of non-flow particle correlations in
McLerran--Venugopalan model \cite{tv,kvn} on top of the ones considered
in \cite{KT}. Those are due to dependence of classical gluonic fields
on nuclear overlap geometry and may contribute to $v_2$
\cite{tv,kvn}. (If taken separately these correlations can not
reproduce the high-$p_T$ behavior of $v_2 (p_T)$ given by RHIC data.) 
We will argue in Sect. 2 that these correlations are higher order in
$\as$ and are therefore parametrically suppressed.

In Sect. 3 we will demonstrate that even in the case of only non-flow
correlations the distribution of particles with respect to
experimentally determined reaction plane averaged over many events is
proportional to $\cos 2 (\phi - \Psi_R)$ in agreement with STAR data
\cite{starsat3}. ($\phi$ and $\Psi_R$ are azimuthal angles of the
particle and reaction plane correspondingly.\footnote{We would like to
point out that the true reaction plane angle $\Phi_R$ used in
\eq{v2diff} is, in principle, different from the reaction plane angle
$\Psi_R$ determined from the flow analysis.}) This distribution is due
to the fact that the reaction plane is also determined from the second
harmonic of the multiplicity distribution. Therefore the $\cos 2 \phi$
shape of the distribution with respect to reaction plane does not
reflect any physics. At the same time the amplitude of the
distribution is given by physical correlations, that is by $2 \, v_2$
\cite{KT}.

In Sect. 4 we show that two-particle correlation functions in our
model are consistent with the data reported by PHENIX
\cite{phenix}. To see that, one has to relax the large rapidity
interval condition employed for simplicity in \cite{KT}, which does
not apply to the PHENIX detector and to flow analysis in general.

We conclude in Sect. 5 by discussing higher cumulant flow
analysis. Higher cumulants analysis was proposed in \cite{cumul} as a
way to reduce the contribution of non-flow effects to $v_2$. We
consider the case when all of standard (2nd cumulant) $v_2$ is due to
non-flow two-particle correlations, similar to \cite{KT}. We then
calculate the higher order cumulants in this model and extract $v_2$
from them. While parametrically the conclusion of \cite{cumul} still
holds, the numerical values of non-flow $v_2$ extracted from the
fourth and higher cumulants for RHIC are not significantly smaller
than $v_2$ from the two-particle correlation function in agreement
with the results of recent STAR analysis \cite{starsat2}.

\section{Different Types of Non-Flow Correlations}

The two-particle multiplicity distribution is given in our model
\cite{KT} by
\be\label{nn2}
\frac{dN}{d^2 k_1 \, dy_1 \, d^2 k_2 \,dy_2} \, = \, 
\frac{dN}{d^2 k_1\, dy_1}\frac{dN}{d^2 k_2\, dy_2} + 
\frac{dN_{corr}}{d^2 k_1\,dy_1\, d^2 k_2\,dy_2}
\ee
with the first term in \eq{nn2} given by a product of two single
particle distributions due to classical gluon fields (disconnected
piece) \cite{claa,kv,yuaa} and the second term corresponding to the
correlations estimated in \cite{KT} (connected piece). Let us make
some parametric estimates in the framework of classical particle
production where the coupling is small $\as (Q_s) \ll 1$ and together
with atomic number $A \gg 1$ it forms a resummation parameter $\as^2
A^{1/3} \sim 1$ \cite{km}. First we note that at the leading order
single particle distribution is
\be\label{par1}
\frac{dN}{d^2 k \, dy} \, \sim \, \frac{S_\perp^A}{\as}
\ee
with $S_\perp^A = \pi R^2$ the cross sectional area of the
nucleus. The gluons produced by classical fields have typical
transverse momentum of the order $k_T \sim Q_s \gg 1/R$ and thus do
not ``know'' of the nuclear overlap geometry. This way the
distribution of these gluon modes in \eq{par1} is azimuthally
symmetric. Therefore the leading contribution of the first term in
\eq{nn2} does not contribute to $v_2$. However, the soft momentum tail
of the single gluon distribution with $k_T \sim 1/R$ does depend on
geometry of the overlap \cite{tv}. If one assumes that large $Q_s$ is
sufficient to keep $\as$ small even for these soft modes one can
perform a calculation of these geometric effects similar to how it was
done in \cite{tv,kvn}. One would obtain the following corrected
version of \eq{par1} \cite{tv}
\be\label{par11}
\frac{dN}{d^2 k \, dy} \, \sim \, \frac{S_\perp^A}{\as} \, \left[1 + 
o\left(\frac{1}{Q_s^2 \, S_\perp^A}\right)\right].
\ee
As $Q_s^2 \sim \as^2 A^{1/3} \sim 1$ the correction in \eq{par11} is
of order $1/\as$. To calculate the contribution of this correction to
elliptic flow observable one can either use \eq{par11} directly in
\eq{v2diff} or use the definition of $v_2$ through two-particle
correlations \cite{wang,ollie}
\be\label{v2corr}
 \left< v_2 \right> \, = \, \sqrt{\left< \, e^{2 i (\phi_1 - \phi_2)}
 \, \right>}.
\ee
In the end one obtains the contribution to $v_2$ of geometric
corrections to classical fields \cite{tv}
\be\label{v2class}
v_2^{geom} \, \sim \, \frac{1}{S_\perp^A}.
\ee
In contrast the second term in \eq{nn2} is parametrically of order \cite{KT}
\be\label{ncorr}
\frac{dN_{corr}}{d^2 k_1\,dy_1\, d^2 k_2\,dy_2} \, \sim \, S_\perp^A.
\ee
This can be understood by combining two classical gluon fields and
letting them interact. This corresponds to squaring \eq{par1},
dropping one power of $S_\perp^A$ (gluons have to be at the same
impact parameter to interact) and multiplying by $\as^2$
(interaction). Using Eqs. (\ref{v2corr}) and (\ref{par1}) the
contribution of correlations in \eq{ncorr} to $v_2$ can be estimated
as \cite{KT}
\be\label{v2nf}
v_2^{non-flow} \, \sim \, \frac{\as}{\sqrt{S_\perp^A}}. 
\ee
To see which contribution to $v_2$ is dominant we observe that
$S_\perp^A \sim A^{2/3} \sim 1/\as^4$ since $\as^2 A^{1/3}
\sim1$. Plugging this into Eqs. (\ref{v2class}) and (\ref{v2nf}) gives 
$v_2^{geom} \sim \as^4$ and $v_2^{non-flow} \sim \as^3$. Therefore
$v_2^{non-flow} \gg v_2^{geom}$ when the coupling is small indicating
that two-particle correlations are parametrically more important for
$v_2$ than geometrical dependence of classical fields. In the actual
RHIC data the saturation scale was estimated to be $Q_s^2
\approx 2$~GeV$^2$ \cite{kln} corresponding to $\as (Q_s) \approx 0.3$ 
which means that geometrical effects of \cite{tv,kvn} may still be
numerically important, though probably only in the soft transverse
momentum region of $k_T \sim 1/R$.

Finally we should mention that HBT correlations are also present in
the model and should in principle be added to the right hand side of
\eq{nn2}. Parametrically HBT correlations are of the order
\be\label{hbt}
\frac{dN^{HBT}}{d^2 k_1\,dy_1\, d^2 k_2\,dy_2} \, \sim \, 
\frac{S_\perp^{A \, 2}}
{\as^2} \, \sim \, \frac{1}{\as^{10}}
\ee
and are much larger than any other correlations mentioned above being
comparable to the uncorrelated piece. (The estimate of \eq{hbt} could
be obtained by squaring \eq{par1}. To derive the HBT correlation
coefficient one has to consider interference of the particle
production amplitudes contributing to the first term in \eq{nn2}
\cite{hbtref}.) However these correlations are important only when
$|\vec{k}_1-
\vec{k}_2| \lsim 1/R$ and could be excluded from experimental flow analysis by
imposing corresponding cuts on particle momenta. The contribution of
HBT correlations to the flow observables was studied in detail in
\cite{hbtflow}.

We should also point out that non-perturbative non-flow correlations
due to jet fragmentation, hadronization and resonance decay are not
included in our model and could be somewhat important numerically in
the data. We assume that their effect is subleading. This assumption
can be illustrated by the following observation. For the elliptic flow
observable $v_2$ we can construct an infrared safe definition using
\eq{v2corr} with $p_T$ weights. Namely one can define the following
observable
\be\label{u2}
\left< u_2 \right>^2 \, = \, \frac{\sum_{i,j} k_i \, k_j \, \cos 2 
(\phi_i - \phi_j)}{\sum_{i,j} k_i \, k_j}
\ee
where the sum goes over all the particles involved in a flow analysis
and $k_i$ are their respective transverse momenta. This observable is
infrared safe, in a sense that it is not sensitive to hadronization of
partons which involves collinear particle splitting and recombination
in the final state \cite{ks}. This implies that fragmentation
functions are not needed in calculation of this observable and the
perturbative calculation similar to the one carried out in \cite{KT}
would be reliable \cite{ks}. The observable in \eq{u2} is different
from the usual definition of $v_2$ with $p_T$ weights
\cite{cumul,pv}. For the saturation minijet model of \cite{KT} the 
observable $\left< u_2 \right>$ can be written as
\be\label{u21}
\left< u_2 \right>^2 \, = \, \frac{1}{\left< k_\perp \right>^2 \, N_{tot}^2} 
\int d^2 k_1 d y_1 d^2 k_2 d y_2  
\frac{d N_{corr}}{d^2 k_1 d y_1 d^2 k_2 d y_2} \, k_1 \, k_2 \, 
\cos 2 (\phi_1 - \phi_2)
\ee
with $\left< k_\perp \right>$ the averaged transverse momentum of the
produced particles and $N_{tot}$ the total multiplicity. The typical
transverse momentum in the particle spectrum generated by the
saturation physics is given by the saturation scale, $\left< k_\perp
\right> \approx Q_s$ \cite{mv,claa,kv,yuaa}. Therefore it would be a 
reasonable approximation to replace $k_1 k_2$ in the integral of
\eq{u21} by $Q_s^2$, which in turn would cancel with $\left< k_\perp 
\right>^2 = Q_s^2$ in the denominator yielding
\be
\left< u_2 \right>^2 \, \approx \, 
\frac{1}{N_{tot}^2} 
\int d^2 k_1 d y_1 d^2 k_2 d y_2 
\frac{d N_{corr}}{d^2 k_1 d y_1 d^2 k_2 d y_2} \,  
\cos 2 (\phi_1 - \phi_2) \, = \, \left< v_2 \right>^2.
\ee
Therefore the elliptic flow observable $v_2$ calculated in the minijet
model of \cite{KT} is, to a good accuracy, equal to the infrared safe
observable $u_2$, and is, therefore, insensitive to parton
fragmentation physics. Perturbative calculation of \cite{KT} is thus
sufficient to give a good estimate for $v_2$ and introduction of
fragmentation functions would not significantly change the result.

\section{Azimuthal Correlations: Reaction Plane Analysis}

\subsection{Simple Example: Autocorrelations}

Let us consider a simple example of scattering events in which $N$
independent particles are produced each having a homogeneous azimuthal
probability distribution $d n / d \phi \equiv (1/N) dN/d\phi = 1/ 2
\pi$. (We set the normalization to $1$ for simplicity.) Let us define
the reaction plane in each event using all $N$ particles according to
the standard flow analysis \cite{pv}. After choosing weights to be
equal to one the reaction plane angle $\Psi_R$ for $v_2$ analysis is
determined by \cite{pv}
\be\label{rpl}
\tan 2 \Psi_R \, = \, \frac{\sum_{i=1}^N \sin 2 \phi_i}{\sum_{j=1}^N
\cos 2 \phi_j}
\ee
where out of two roots between $0$ and $2 \pi$ one has to take the one
with same signs for $\cos 2 \Psi_R$ and $\sum_{j=1}^N
\cos 2 \phi_j$. Here the ``reaction plane angle'' $\Psi_R$ of course has 
nothing to do with the real reaction plane since all particles are
assumed to be produced independent of geometry and of each other. Our
goal is to calculate azimuthal distribution of one of the particles
with respect to reaction plane determined by \eq{rpl}. Of course the
correlation of each of the particles with the reaction plane in
\eq{rpl} is due to our inclusion of this particle in the definition of
the reaction plane and is not physical. It is, nevertheless,
instructive to see what the azimuthal shape of these autocorrelations
would be.

The distribution of particles with respect to the reaction plane is
defined as
\ben
\frac{d n}{d \phi_1 \ d \Psi_R} \, = \, \int_0^{2 \pi} 
d \phi_2 \ldots d \phi_N  
\frac{d n}{d \phi_1 \, d \phi_2 \ldots d \phi_N} \, \delta\left
(\tan 2 \Psi_R - \frac{\sum_{i=1}^N \sin 2 \phi_i}{\sum_{j=1}^N
\cos 2 \phi_j}\right)
\een
\be\label{dist1}
\times \, \frac{1}{\cos^2 2 \Psi_R} \, \theta\left( \cos 2 \Psi_R \sum_{k=1}^N
\cos 2 \phi_k\right)
\ee
with the distribution function for $N$ independent particles
\be
\frac{d n}{d \phi_1 \, d \phi_2 \ldots d \phi_N} \, = \, \frac{d n}{d \phi_1} 
\, \frac{d n}{d \phi_2} \, \ldots \,  \frac{d n}{d \phi_N} \, = \, 
\frac{1}{(2 \pi)^N}.
\ee
For simplicity we will put $\Psi_R = 0$ in what follows. Representing
$\delta$- and $\theta$- functions as integrals we rewrite \eq{dist1}
as
\ben
\frac{d n}{d \phi_1 \ d \Psi_R} \Bigg|_{\Psi_R=0} \, = \, 
\int_{-\infty}^\infty 
\frac{d \xi}{2 \pi} \, \frac{d \eta}{2 \pi i} \, \frac{1}{\eta - i \epsilon} 
\, \int_0^{2 \pi} \frac{d \phi_2 \ldots d \phi_N}{(2 \pi)^N} \, \sum_{j=1}^N
\cos 2 \phi_j 
\een
\be\label{aut1}
\times \, e^{- i \, \xi \, \sum_{i=1}^N \sin 2 \phi_i + i \, \eta \, 
\sum_{k=1}^N \cos 2 \phi_k}.
\ee
After integration over angles the expression becomes 
\ben
\frac{d n}{d \phi_1 \ d \Psi_R} \Bigg|_{\Psi_R=0} \, = \, \frac{1}{2 \pi} 
\, \int_{-\infty}^\infty \frac{d \xi}{2 \pi} \, \frac{d \eta}{2 \pi i} \, 
\frac{1}{\eta - i \epsilon} \, e^{- i \, \xi \, \sin 2 \phi_1 + i \, \eta \, 
\cos 2 \phi_1} \, \left\{ \cos 2 \phi_1 \, \left[ J_0 (\sqrt{\xi^2 + \eta^2})
\right]^{N-1} \right.
\een
\be\label{aut2}
\left. + \, (N-2) \, \frac{i \, \eta}{\sqrt{\xi^2 + \eta^2}} \, 
J_1 (\sqrt{\xi^2 + \eta^2}) \, 
 \left[ J_0 (\sqrt{\xi^2 + \eta^2}) \right]^{N-2} \right\},
\ee
where the first term in the curly brackets of \eq{aut2} corresponds to
$j=1$ term in the sum of \eq{aut1}, while the second term includes the
rest of the sum. To evaluate \eq{aut2} in the $N \rightarrow \infty$
limit let us note that since $J_0 (\sqrt{\xi^2 + \eta^2}) < 1$ for all
real non-zero $\sqrt{\xi^2 + \eta^2}$ and $J_0 (0) = 1$ the integrals
in \eq{aut2} are dominated by small values of $\xi$ and
$\eta$. Expanding the Bessel function we write
\be\label{aut3}
\left[ J_0 (\sqrt{\xi^2 + \eta^2}) \right]^{N-1} 
\Bigg|_{N \rightarrow \infty} 
\, \approx \, \exp \left[ - \frac{N (\xi^2 + \eta^2)}{4} \right]. 
\ee
The second term in the brackets of \eq{aut2} becomes
\be\label{aut4}
\frac{1}{2 \pi} 
\, \int_{-\infty}^\infty \frac{d \xi}{2 \pi} \, \frac{d \eta}{2 \pi} 
\, \frac{N}{2} \, \exp \left[ - \frac{N (\xi^2 + \eta^2)}{4} \right] 
\, = \, \frac{1}{(2 \pi)^2} 
\ee
where we put the exponent of \eq{aut2} to one in our large-$N$
approximation. Due to \eq{aut3} the integral in \eq{aut2} is dominated
by $\xi \sim \eta \sim 1/\sqrt{N}$. Therefore expansion of exponent in
\eq{aut2} beyond $0$th order gives $o(1/N)$ subleading
corrections.

The first term in \eq{aut2} gives
\be\label{aut5}
\frac{1}{2 \pi} \, \cos 2 \phi_1 \, 
\, \int_{-\infty}^\infty \frac{d \xi}{2 \pi} \, \frac{d \eta}{2 \pi i} \, 
\frac{1}{\eta - i \epsilon} \, \exp \left[ - \frac{N (\xi^2 + \eta^2)}{4} 
\right] \, = \, \frac{1}{(2 \pi)^2} \, \cos 2 \phi_1 \, \sqrt{\frac{\pi}{N}}
\ee
where we have used
\ben
\frac{1}{\eta - i \epsilon} \, = \, \mbox{P.V.} \frac{1}{\eta} \, 
+ \, \pi \, i \, \delta (\eta).
\een
Combining Eqs. (\ref{aut4}) and (\ref{aut5}) and inserting non-zero
$\Psi_R$ we end up with
\be\label{auto}
\frac{d n}{d \phi \ d \Psi_R} \, = \, \frac{1}{(2 \pi)^2} \, \left( 1 + 
\sqrt{\frac{\pi}{N}} \, \cos 2 (\phi - \Psi_R) \right)
\ee
up to $o(1/N)$ corrections.

We can draw the following general conclusion from \eq{auto}. If a
particle is included in the reaction plane angle definition from the
$n$--th harmonic it is correlated to the reaction plane with the shape
of the correlations' distribution in the large multiplicity limit
given by $\cos n (\phi - \Psi_R)$. Note that (auto)correlations
leading to $\cos 2 \phi$ distribution of \eq{auto} are non-flow
correlations by their definition.

\subsection{Correlations with Respect to Reaction Plane}

Now we are going to generalize the simple model of the previous
subsection to the case of some weak pairwise physical correlations
between the particles, similar to the correlations considered in
\cite{KT}. Namely let us consider the case where the particle number $1$,
which correlations to the reaction plane will be studied, is excluded
from reaction plane definition but may have some physical correlations
with the particles $2, \ldots, N$ contributing to reaction plane
determination. We thus remove autocorrelation of particle number $1$
with reaction plane in accordance with the standard method of flow
analysis \cite{pv}.  However, particle $1$ can now be physically
correlated to, say, particle $2$. The latter is included in reaction
plane definition and is therefore autocorrelated to the reaction plane
with $\cos 2 \phi$ distribution of \eq{auto}. Our claim here is that
it is this $\cos 2 \phi$ autocorrelations distribution, and not the
physical correlations, which will be dominant in determining the
azimuthal shape of the correlations of particle $1$ to the reaction
plane. In what follows we will keep in mind the cuts commonly employed
in flow analysis, which were recently used by STAR \cite{starsat3}. In
\cite{starsat3} only particles with $p_T < 2$~GeV were included in 
the reaction plane definition (corresponding to our particles $2,
\ldots , N$) and correlations of particles with $p_T > 2$~GeV 
(corresponding to our particle $1$) to this reaction plane were
studied.

We start by defining a correlated two-particle distribution given for
example by minijet correlation discussed in \cite{KT} and normalized
according to
\be
\frac{d n_{corr}}{d \phi_1 \, d \phi_2} \, = \, \frac{1}{N (N-1)} \, 
\frac{d N_{corr}}{d \phi_1 \, d \phi_2}. 
\ee
The distribution of $N$ particles now becomes
\be\label{dcorr}
\frac{d n}{d \phi_1 \, d \phi_2 \ldots d \phi_N} \, = \, 
\frac{d n}{d \phi_1} \, \frac{d n}{d \phi_2} \, \ldots \,  
\frac{d n}{d \phi_N} + (N - 1) \, \frac{d n_{corr}}{d \phi_1 \, 
d \phi_2} \, \frac{d n}{d \phi_3} \, \ldots \, \frac{d n}{d \phi_N} +
\ldots
\ee
where the factor of $N-1$ accounts for all possible pairwise
correlations of particle $1$ with particles $2, \ldots , N$. In
\eq{dcorr} we omit possible pairwise correlation of particles $2,
\ldots , N$ with each other which are of the same order as the
correlations shown in \eq{dcorr} but do not contribute to the
azimuthal asymmetry of the distribution of particle $1$ with respect
to the reaction plane, giving only an additive small constant. The
correlations of more than one pair of particles are neglected in
\eq{dcorr} since they are suppressed by higher powers of $\as$ \cite{KT}.

To determine the distribution of particle $1$ with respect to reaction
plane we have to substitute distribution of \eq{dcorr} into
\eq{dist1}. Evaluation of the first term in the resulting expression 
could be easily done repeating the steps which led to the first term
in \eq{auto}. One thus obtains
\ben
\frac{d n}{d \phi_1 \ d \Psi_R} \Bigg|_{\Psi_R=0} \, \approx \, 
\frac{1}{(2 \pi)^2} \, + \, (N-1) \, \int_{-\infty}^\infty 
\frac{d \xi}{2 \pi} \, \frac{d \eta}{2 \pi i} \, \frac{1}{\eta - 
i \epsilon} \, \int_0^{2 \pi} \frac{d \phi_2 \ldots d \phi_N}{(2
\pi)^{N-2}} \, \, \frac{d n_{corr}}{d \phi_1 \, 
d \phi_2}
\een
\be\label{corr1}
\times  \, \sum_{j=2}^N \cos 2 \phi_j \, e^{- i \, \xi \, 
\sum_{i=2}^N \sin 2 \phi_i + i \, \eta \, 
\sum_{k=2}^N \cos 2 \phi_k}.
\ee
We are interested only in $j=2$ term in \eq{corr1} since it comes with
$\cos 2\phi_2$ and can give non-trivial correlations with the reaction
plane. In all the other terms in the sum of cosines the integration
over $\phi_2$ would give an additive constant small compared to the
first term on the right hand side of \eq{corr1}. Assuming that $d
n_{corr} / d \phi_1 \, d \phi_2$ is an even function of $\phi_1 -
\phi_2$ only (which is, probably, true for all perturbative 
spin-independent pairwise correlations and is certainly true for
correlations considered in \cite{KT}) we write
\ben
\int_0^{2 \pi} d \phi_2 \, \frac{d n_{corr}}{d \phi_1 \, 
d \phi_2} \, \cos 2 \phi_2 \, = \, \cos 2 \phi_1 \, \int_0^{2 \pi} d
\phi_2 \, \frac{d n_{corr}}{d \phi_1 \, d \phi_2} \, \cos 2 (\phi_1 - 
\phi_2) 
\een
\be\label{corr2}
= \, \cos 2 \phi_1 \, 2 \pi \, v_2 (1) \, v_2 (2) \, 
\frac{dn}{d\phi_1} \, \frac{dn}{d\phi_2} \, = \, \cos 2 \phi_1 \, 
\frac{1}{2 \pi} \, v_2 (1) \, v_2 (2),
\ee
where we have used the definition of flow observable from \eq{v2corr}
\cite{wang,ollie} and substituted $dn/d\phi = 1/2\pi$. $v_2 (1)$ and 
$v_2 (2)$ are elliptic flow variables for the particles $1$ and $2$
correspondingly. Let us now consider a specific example of flow
analysis when the particle $1$ represents all particles with fixed
value of transverse momentum $p_T$ while particle $2$ represents
particles in a broad transverse momentum range which go into reaction
plane definition (e.g. all particles with $p_T < 2$~GeV as in STAR
analysis \cite{starsat3}). Then $v_2 (1)$ would correspond to
differential elliptic flow $v_2 (p_T, B)$, while $v_2 (2)$ would be
the averaged elliptic flow $v_2 (B)$, where $B$ is the impact
parameter of the collision. With the help of \eq{corr2}, \eq{corr1}
becomes
\ben
\frac{d n}{d \phi_1 \ d \Psi_R} \Bigg|_{\Psi_R=0} \,\approx \, 
\frac{1}{(2 \pi)^2} \, + \, \cos 2 \phi_1 \, 
\frac{1}{2 \pi} \, (N-1) \, v_2 (p_T, B) \, v_2 (B) 
\een
\be\label{corr3}
\times \, \int_{-\infty}^\infty 
\frac{d \xi}{2 \pi} \, \frac{d \eta}{2 \pi i} \, \frac{1}{\eta - 
i \epsilon} \, \left[ J_0 (\sqrt{\xi^2 + \eta^2})\right]^{N-2}, 
\ee
which can be rewritten in the large-$N$ approximation as
\ben
\frac{d n}{d \phi_1 \ d \Psi_R} \Bigg|_{\Psi_R=0} \,\approx \, 
\frac{1}{(2 \pi)^2} \, + \, \cos 2 \phi_1 \, 
\frac{1}{2 \pi} \, N \, v_2 (p_T, B) \, v_2 (B) 
\een
\be\label{corr4}
\times \, \int_{-\infty}^\infty 
\frac{d \xi}{2 \pi} \, \frac{d \eta}{2 \pi i} \, \frac{1}{\eta - 
i \epsilon} \, \exp \left[ - \frac{N (\xi^2 + \eta^2)}{4} \right]. 
\ee
Performing the integrations in \eq{corr4} similar to how it was done
in obtaining \eq{aut5} and reintroducing non-zero $\Psi_R$ we write
\be\label{corr5}
\frac{d n}{d \phi_{p_T} \ d \Psi_R} \, \approx \, 
\frac{1}{(2 \pi)^2} \, \left[ 1 +  \, \sqrt{\pi N} \, v_2 (p_T, B) 
\, v_2 (B) \, \cos 2 (\phi_{p_T} - \Psi_R) \right]. 
\ee
\eq{corr5} represents the main point of this Section: the shape of 
the correlations of particles with respect to the reaction plane
determined by the standard elliptic flow analysis is proportional to
$\cos 2 (\phi_{p_T} - \Psi_R)$ independent of the physical nature of
these correlations. Therefore particle distribution with respect to
reaction plane carries no information about the azimuthal shape of
physical correlations.

Nevertheless, the coefficient in front of the cosine in \eq{corr5} is
given by the strength of physical correlations. To understand the
nature of the coefficient in \eq{corr5} let us calculate reaction
plane resolution in our model. If we divide all $N-1$ particles
defining the reaction plane into two subgroups of (roughly) $N/2$
particles each and determine the reaction plane angles in each
subgroup independently ($\Psi_1$ and $\Psi_2$), the reaction plane
resolution would be given by \cite{pv}
\be\label{rpl1}
\Delta \, = \, \sqrt{2 \left< \cos 2 (\Psi_1 - \Psi_2)\right>}
\ee
where the averaging is taken over many same--multiplicity events. To
determine the strength of the reaction plane correlations in our model
we note that the sub-event planes are correlated in it only due to
pairwise correlations between the particles. One particle from plane
$\Psi_1$ is correlated to plane $\Psi_2$ via the distribution in
\eq{corr5}. The same particle is correlated to its own plane $\Psi_1$
via autocorrelations of \eq{auto}. To determine correlations between
the two planes introduced by this one particle we have to multiply the
two distributions and average over all angles and momenta of the
particle $\phi$. The total correlations would be obtained by taking
this correlation due to one particle to the $N/2$ power to account for
all $N/2$ particles in the sub-event plane definition. Expanding the
resulting expression up to the lowest order in correlations we end up
with
\be\label{rpl2}
\frac{d n}{d \Psi_1 \, d \Psi_2} \, \propto \, 1 + \frac{\pi}{4} \, 
N \, v_2 (B)^2 \, \cos 2 (\Psi_1 - \Psi_2).
\ee
Using Eqs. (\ref{rpl1}) and (\ref{rpl2}) we can determine the reaction 
plane resolution as
\be\label{rpl3}
\Delta \, = \, \frac{\sqrt{\pi \, N}}{2} \, v_2 (B),
\ee
which is in agreement with \cite{pv,vz}. With the help of \eq{rpl3},
\eq{corr5} can be rewritten as
\be\label{correl}
\frac{d n}{d \phi_{p_T} \ d \Psi_R} \, \approx \, 
\frac{1}{(2 \pi)^2} \, \left[ 1 +  \, 2 \, v_2 (p_T, B) 
\, \Delta \, \cos 2 (\phi_{p_T} - \Psi_R) \right]. 
\ee
Now we can see that, independent of the nature of correlations
contributing to $v_2$, particle distribution with respect to reaction
plane has a $\cos 2 \phi$ shape with the amplitude given by
differential elliptic flow observable $v_2 (p_T, B)$ times the event
plane resolution. In a recent paper by STAR \cite{starsat3} the data
on particle distribution with respect to reaction plane was reported,
which was fitted rather well with the ansatz of \eq{correl}
\cite{starsat3}. \eq{correl} tells us that since our model describes
the data on $v_2 (p_T, B)$ rather well \cite{KT} it also agrees with
the recent data on correlations with respect to reaction plane
reported by STAR
\cite{starsat3}.

\section{Azimuthal Correlations: Correlation Functions}


We saw in the previous section that the reaction plane analysis cannot
determine the physical shape of azimuthal particle correlations.  In
this section we consider another method of azimuthal correlations
analysis -- the correlation function method. We believe that this
method correctly determines the shape of azimuthal correlations. The
two-particle azimuthal correlation function is defined by
\cite{phenix}
\be\label{defcorr}
C(\Delta \phi)\,=\, \frac{dN_{real}/d \Delta \phi}
{dN_{mixed}/d \Delta \phi} \, \frac{N_{mixed}}{N_{real}},
\ee
where $dN_{real}/d\Delta\phi$ is the number of particle pairs observed
in the same event with a given azimuthal opening angle $\Delta\phi$
and $dN_{mixed}/d\Delta\phi$ is the number of particle pairs selected
from two different events with the same azimuthal opening
angle. $N_{real}$ and $N_{mixed}$ are total numbers of pairs in the
same and in different events correspondingly. Only events with the
same multiplicity are selected for the analysis of $C(\Delta\phi)$. In
this section we will calculate the contribution of two-particle
correlations in our minijet model \cite{KT} to the azimuthal
correlation function.

The number of particle pairs produced in the same event with a given
azimuthal opening angle $\Delta\phi$ can be obtained by integrating
\eq{nn2}
\be\label{real1}
\frac{dN_{real}}{d\Delta\phi}\,=\, 
2 \pi \, \int\, d\, k_1\, k_1 \, dk_2\, k_2\, dy_1\,
dy_2\, \frac{dN}{d^2 k_1\, dy_1\, d^2 k_2\, dy_2},
\ee
while the total number of pairs is
\be\label{real2}
N_{real} \, = \, \int\, d^2 \, k_1\, d^2 k_2\, dy_1\,
dy_2\, \frac{dN}{d^2 k_1\, dy_1\, d^2 k_2\, dy_2}.
\ee
Particle pair where particles are selected from two different events
obviously have no physical correlations. Therefore their azimuthal 
angle distribution is trivial and is given by 
\be\label{nmixed}
\frac{1}{N_{mixed}} \, \frac{dN_{mixed}}{d\Delta\phi}\,=\, \frac{1}{2\pi}.
\ee
Combining Eqs. (\ref{real1}), (\ref{real2}), (\ref{nmixed}) and
(\ref{nn2}) with \eq{defcorr} yields 
\be\label{corrall}
C(\Delta \phi)\,=\, \frac{(2 \pi)^2 \, \int  
d\, k_1\, k_1 \, dk_2\, k_2\, dy_1\,
dy_2\, \left(\frac{dN}{d^2 k_1\, dy_1}
\frac{dN}{d^2 k_2\, dy_2} + \frac{dN_{corr}}{d^2 k_1\,dy_1\, 
d^2 k_2\,dy_2}\right)}{ \int\, d^2 \, k_1\, d^2 k_2\, dy_1\,
dy_2\, \left( \frac{dN}{d^2 k_1\, dy_1}\frac{dN}{d^2 k_2\,
dy_2} +
\frac{dN_{corr}}{d^2 k_1\,dy_1\, d^2 k_2\,dy_2}\right)}. 
\ee
To calculate the correlation function we, therefore, need to know
single- and double- inclusive particle distributions.

We calculated the two-particle multiplicity distribution of \eq{nn2}
in our previous work \cite{KT} assuming that there is a large rapidity
interval $|y_1-y_2|\, \sim \, 1/\as \, \gg 1$ between the correlated
particles. It turns out that this approximation preserves all
important ingredients of the model as long as we are not concerned
with the rapidity dependence of elliptic flow and with the azimuthal
angular distributions. At the same time it considerably simplifies
calculations allowing for a simple interpretation of the final
result. However, the large rapidity interval assumption does not hold
in the actual flow analyses. In order to properly describe data one
has to relax the $|y_1-y_2| \gg 1$ condition. Below we intend to
compare our model's predictions to the correlation function data
reported by PHENIX \cite{phenix}. The (pseudo)rapidity acceptance of
the PHENIX detector is limited to the interval of $0.7$ units in the
central rapidity region and the $|y_1-y_2| \gg 1$ condition certainly
does not hold there.

Single particle distribution was written in our model \cite{KT} as
\be\label{DISTR1}
\frac{dN}{d^2 k_1\,dy_1}\,=\,
\frac{2 \, \as\,}{C_F \, S_\bot}\,\frac{1}{\un k_1^2}\,\int\,
d^2 q_1\frac{dxG_A}{d\un q_1^2}\,\frac{dxG_A}{d(\un k_1- \un q_1)^2}.
\ee
Relaxing $|y_1-y_2| \gg 1$ condition to $|y_1-y_2| \sim 1$ yields a
more general expression for the two-particle distribution
\ben
\frac{dN_{corr}}{d^2 k_1\,dy_1\, d^2 k_2\,dy_2}\,= \, 
\frac{N_c \, \as^2}{\pi^2 \, C_F \, S_\bot}\,\int\,
\frac{d^2 q_1}{\un q_1^2}\,\int\, \frac{d^2 q_2}{\un q_2^2}\,
 \delta^2 (\un q_1+\un q_2-\un k_1-\un k_2)
\een
\be\label{DISTR2}
\times \, \frac{d xG_A}{d\un q_1^2}\,\frac{d xG_A}{d\un q_2^2}\,
\mathcal{A}(\un q_1,\,\un q_2,\,\un k_1, \un k_2, y_1-y_2),
\ee
where $\mathcal{A}(\un q_1,\,\un q_2,\,\un k_1, \un k_2, y_1-y_2)$ is
the two-to-four particles amplitude in the quasi-multi-Regge
kinematics ($|y_1-y_2| \sim 1$).  $\mathcal{A}$ was found in an
impressive calculation performed in \cite{leostr}. We refer the reader
to the reference \cite{leostr} for an explicit expression for
$\mathcal{A}(\un q_1,\,\un q_2,\,\un k_1, \un k_2, y_1-y_2)$. In the
leading logarithmic approximation $|y_1-y_2| \sim 1/\as \gg 1$ this
amplitude reduces to
\be
\mathcal{A}(\un q_1,\,\un q_2,\,\un k_1, 
\un k_2, y_1-y_2)\,=\,\frac{\un q_1^2\, \un q_2^2}{\un k_1^2\, \un k_2^2}.
\ee
This expression for $\mathcal{A}$ was used in our model before
\cite{KT} yielding only back-to-back correlations between the
particles. It was argued in \cite{leostr} that inclusion of the
next-to-leading logarithmic corrections is essential for understanding
the angular distribution of the correlation function. In particular,
on top of the back-to-back correlations, they introduce correlations
at small angles $\Delta \phi$ between the particles in a pair
\cite{leostr}.

The unintegrated gluon distribution due to the quasi-classical 
non-Abelian Weisz\"acker-Williams (WW) field $\un A^{WW}(\un z)$
of the nucleus is given by \cite{claa,km}
\begin{eqnarray}\label{uiglue}
\frac{dxG_A(x,\un q^2)}{d\un q^2} &=&
\frac{2}{(2\pi)^2}\,\int\, d^2\un z \,e^{-i\un z\cdot\un q}\,
\int\, d^2\un b \,\mathrm{Tr}\,\langle \un A^{WW}(\un 0)\,\un A^{WW}(\un
z)\rangle\nonumber\\
&=&
\frac{2}{\pi (2\pi)^2}\,\int\, d^2\un z\, e^{-i\un z\cdot\un q}\, 
\frac{S_\bot C_F}{\as\,\un z^2}\left(
1-e^{-\frac{1}{4}\un z^2 Q^{cl\,2}_s}\right),\label{WW}
\end{eqnarray}
where $\un b$ is the gluon's impact parameter (which we can trivially
integrate over in a cylindrical nucleus case considered here) and
$Q_s^{cl}$ is a certain scale at which nonlinear nature of the gluon
field becomes evident.  In \cite{KT} we argued that one has to include
quantum evolution effects in quasi-classical formulae in order to
describe some important features of the particle spectrum. In
particular we took into account the fact that evolved gluon
distribution scales as a function of $p_T/Q_s$ only, where $Q_s$ is a
saturation scale which emerges from the solution to the nonlinear
evolution equation \cite{bk}. We argued in \cite{KT} that evolution
effects can be mimicked by writing the argument of the Glauber
exponent of \eq{uiglue} in the form
\be
-\frac{1}{4}\,\un z^2 Q^{cl\,2}_s(\un z)\,=\,
- (\xi^2 Q_s^2/ 4 \un q^2) \, \frac{\ln\frac{Q_s}{\xi \, \Lambda}}{\ln
(Q_s/\Lambda)}
\ee
where $\xi\,=\,q\,z$ and $\Lambda$ is the infrared cutoff. Thus, after
integration over angles in \eq{uiglue} we obtain
\be\label{glue}
\frac{dxG_A(x,\un q^2)}{d\un q^2}\,=\,
\frac{1}{\pi^2}\,\frac{S_\bot C_F}{\alpha_s}
\int_0^{Q_s/\Lambda}\, \frac{d\xi}{\xi} \, J_0(\xi)\, 
\frac{S_\bot C_F}{\as\,\un z^2}\left(
1-e^{- (\xi^2 Q_s^2/ 4 \un q^2) \, \frac{\ln\frac{Q_s}{\xi \, \Lambda}}{\ln
(Q_s/\Lambda)}}\right)
\ee
\begin{figure}
\begin{center}
\begin{tabular}{cc}
\epsfig{file=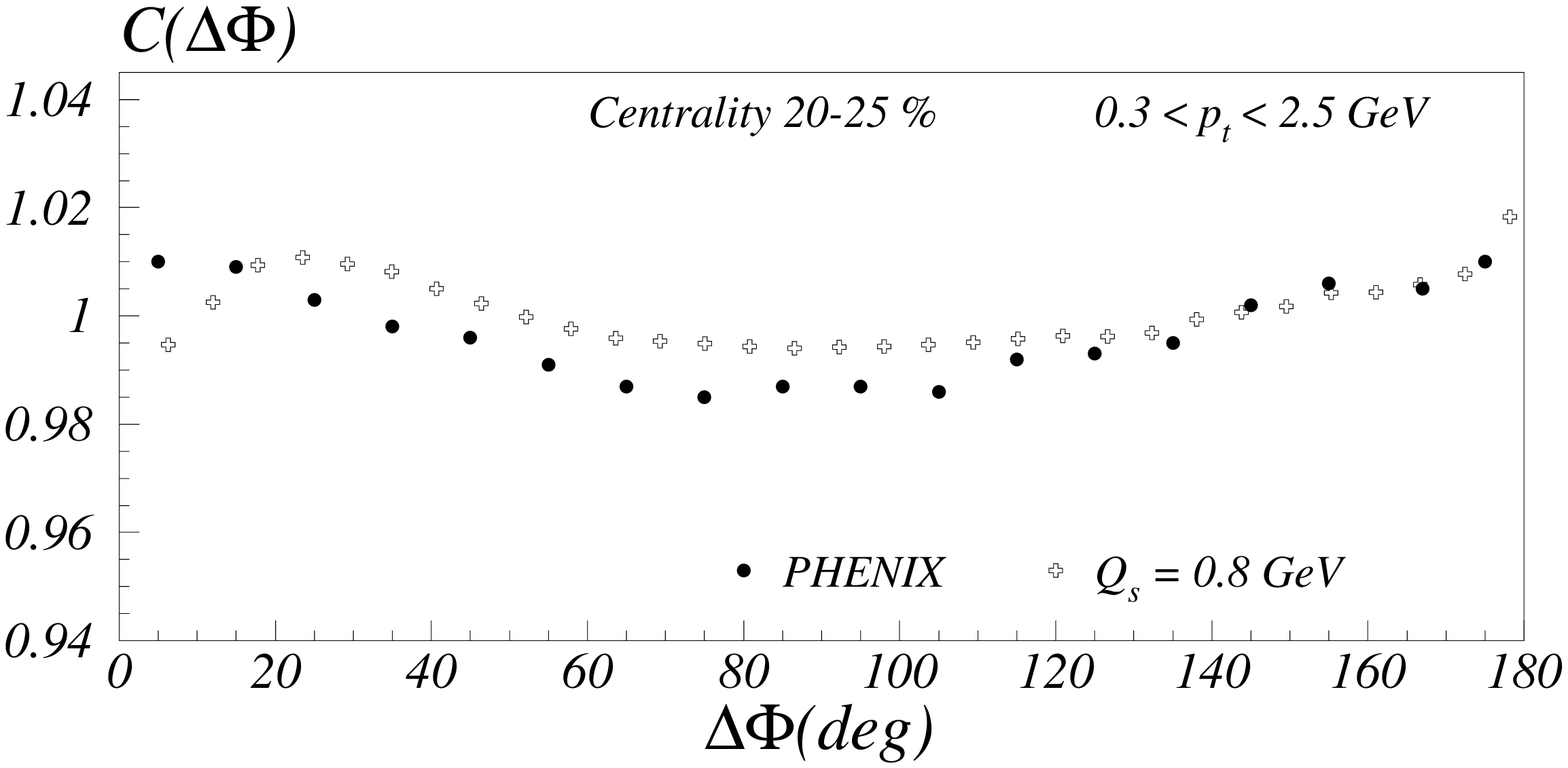, width=8.cm}
&
\epsfig{file=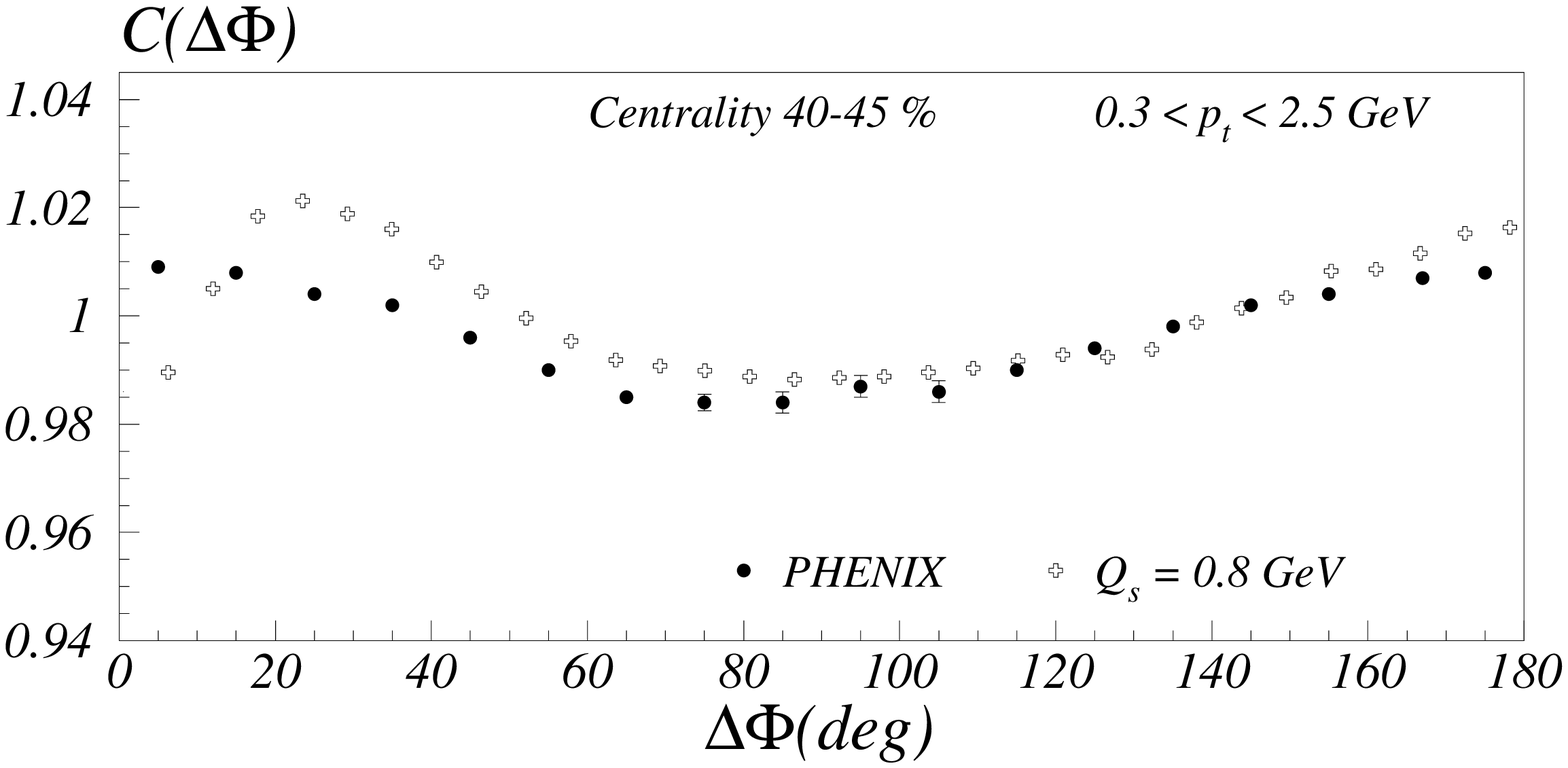, width=8.cm}\\
\epsfig{file=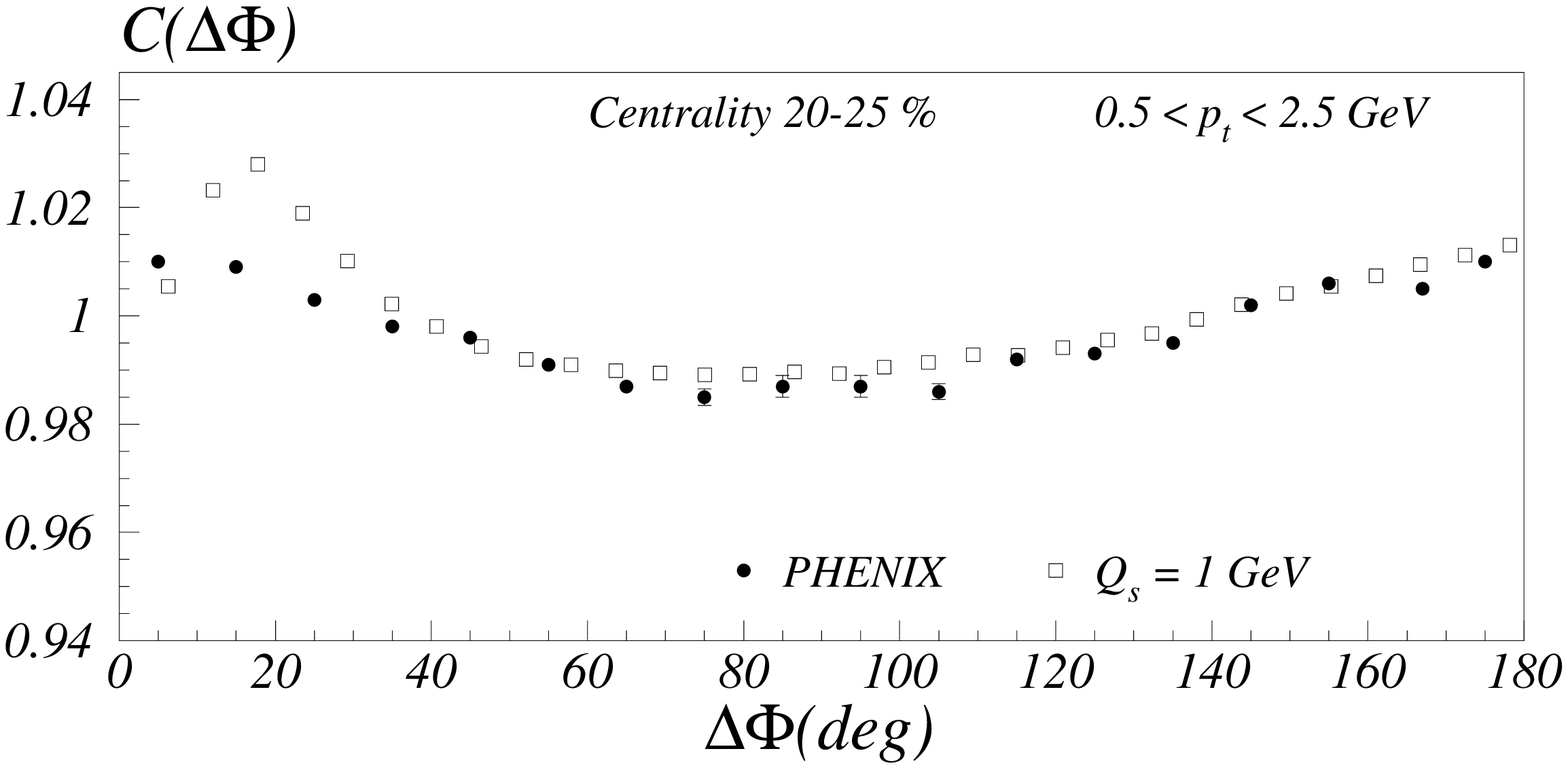, width=8.cm}
&
\epsfig{file=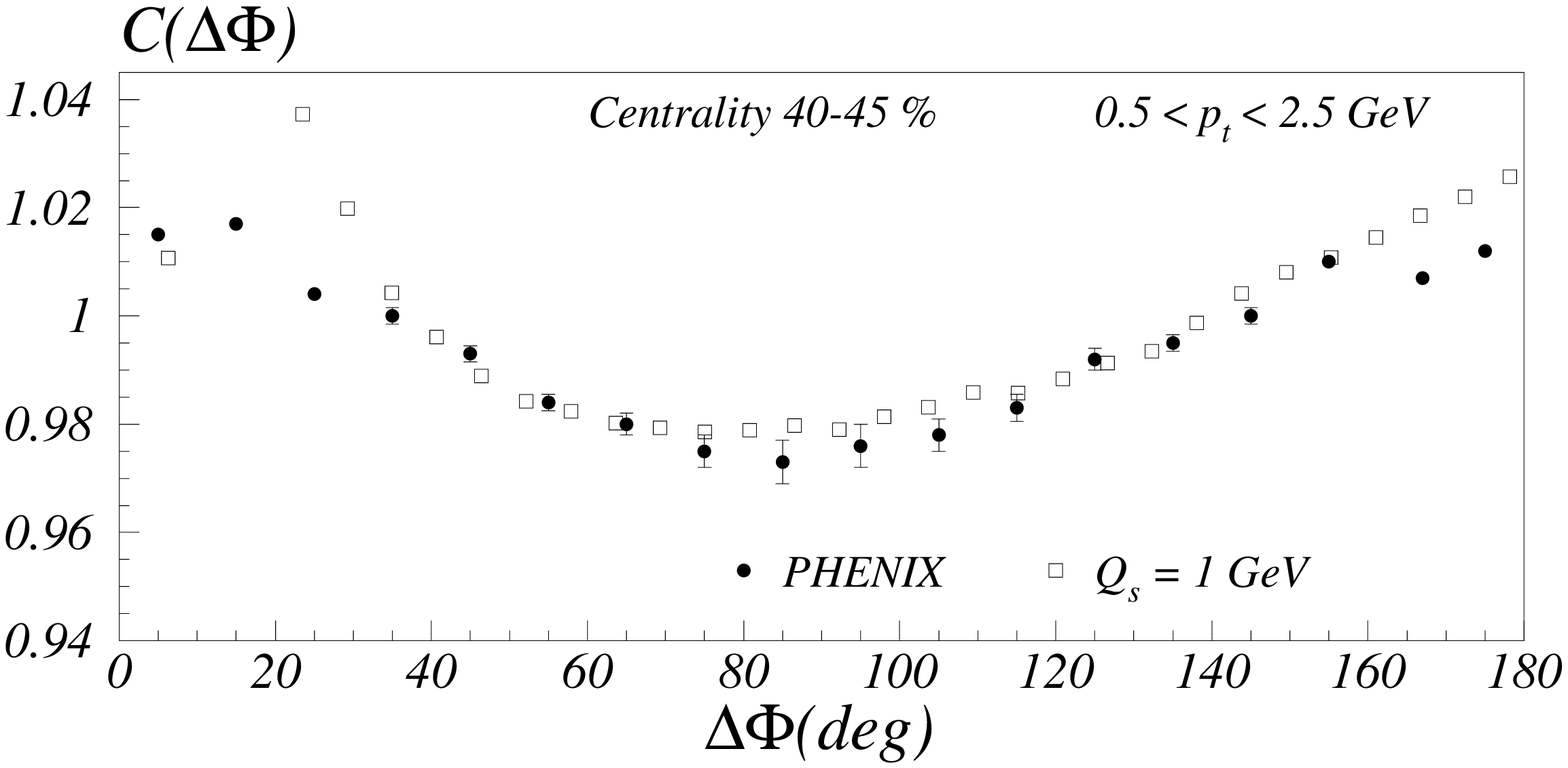, width=8.cm}
\end{tabular}
\end{center}
\caption{\sl Two-particle azimuthal angle correlation function given by 
  our model compared to PHENIX $\sqrt{s}=130$~GeV data for different
  cuts in the transverse momentum and centralities. We used the
  following values of parameters: $\Lambda\,=\,0.15$ GeV, $A=197$,
  $\alpha_s=0.3$.}
\label{fig:corr}
\end{figure}
Formulae (\ref{corrall}), (\ref{DISTR1}), (\ref{DISTR2}), and
(\ref{glue}), together with the expression for $\mathcal{A}$ given in
\cite{leostr} comprise all necessary theoretical information for
calculation of the correlation function $C(\Delta\phi)$. In order to
compare our correlation function to PHENIX data \cite{phenix} we have
to integrate in \eq{corrall} over $y_1$ and $y_2$ in the rapidity
interval from $-0.35$ to $+0.35$ (PHENIX detector acceptance) and over
$k_1$ and $k_2$ in the transverse momentum intervals specified by the
data analysis of \cite{phenix}. The two-particle distribution of
\eq{DISTR2} with $\mathcal{A}$ from \cite{leostr} has a collinear
singularity at small opening angles $\Delta \phi$ due to the
contribution of $1\rightarrow2$ gluon splitting \cite{leostr}. It is
possible that the corresponding increase of the correlation function
would be reduced after inclusion of higher order perturbative
corrections \cite{ciafaloni}. The small angle particle correlations
also receive contributions from HBT correlations, resonance decays and
hadronization functions. While some of these correlations are excluded
by appropriate cuts in the experimental flow analysis, some might
still remain. To model the effect of these phenomena on the
correlation function in our model we have imposed a lower cutoff on
the invariant mass of the two particles produced in \eq{DISTR2}, with
the value of the cut obtained from comparison to the data
\cite{phenix}. The best fit was obtained by requiring that $(k_1 + k_2)^2
\,=\, 2\,(k_1\, k_2 \cosh(\Delta y)\, - \, {\underline k_1} \cdot
{\underline k_2})\, >\, 0.07$~GeV$^2$.

Results of our calculations are presented in
\fig{fig:corr}. The values of parameters that were used are specified in the 
caption of \fig{fig:corr}. Integrations over ${\un q}_1$ and ${\un
q}_2$ in \eq{DISTR2} were cut off in the infrared limit of ${\un q}_1=
0$ and ${\un q}_2= 0$ by $2 \Lambda$.  The value of the saturation
scale giving the best fit for $0.3 < p_T < 2.5$~GeV data is slightly
lower than the saturation scale giving the best fit for $0.5 < p_T <
2.5$~GeV data ($Q_s = 0.8$~GeV and $Q_s = 1$~GeV
correspondingly). This slight discrepancy is probably due to the fact
that non-perturbative (soft) effects become more important for lower
$p_T$ data ($0.3 < p_T < 2.5$~GeV) which could be effectively mimicked
by reducing the saturation scale. The fall-off of our correlation
function at the very small angles is an artifact of the applied
invariant mass cut. More research is needed to completely
quantitatively understand the correlation function at small opening
angles $\Delta\phi$. Overall we observe a good agreement of our model
with experimental data reported by PHENIX \cite{phenix}.


\section{Higher Order Cumulants}

A new method of elliptic flow analysis was recently proposed in
\cite{cumul}. It was suggested to measure flow using not only 
two-particle correlations of \eq{v2corr} \cite{wang, ollie} but also
four-, six- and other higher order correlation functions. The flow
observable can be extracted from cumulants formed out of these
correlation functions. For instance the fourth order cumulant for
elliptic flow is defined as
\cite{cumul}
\be\label{c4}
c_2 \{ 4 \} \, \equiv \, \left<\left< e^{2 i (\phi_1 + \phi_2 - \phi_3
- \phi_4)}\right>\right> \, \equiv \, \left< e^{2 i (\phi_1 + \phi_2 -
\phi_3 - \phi_4)}\right> - 2 \, \left< e^{2 i (\phi_1 - \phi_3)}\right>
\ee
and in case of only flow correlations is equal to \cite{cumul}
\be\label{c4fl}
c_2 \{ 4 \} \, = \, - v_2^4 . 
\ee
The advantage of the higher order cumulant analysis is in the fact
that, as argued in \cite{cumul}, if flow is larger than non-flow
correlations, the contribution of the latter to $v_2$ extracted from
higher order cumulants is suppressed by powers of particle
multiplicity. For instance, if $M$ particles are emitted in a
collision it can be easily shown that the contribution of non-flow
correlations to $v_2$ from \eq{v2corr} scales as $1/\sqrt{M}$. At the
same time a similar analysis shows that non-flow $v_2$ extracted from
the fourth order cumulant of \eq{c4fl} scales as $1/M^{3/4}$, i.e., it
is suppressed by an extra factor of $1/M^{1/4}$. Corresponding data
analysis based on Eqs. (\ref{c4}) and (\ref{c4fl}) has been carried
out at STAR \cite{starsat2}.

In saturation particle production models one should use the number of
participants $N_{part} \sim S_\perp^A Q_s^2$ instead of $M$
\cite{kln,KT}. For $\sqrt{s} = 130$~GeV at RHIC the number of
participants for the most central collisions is $N_{part} (B=0) = 344$
\cite{kln} which gives a suppression factor $1/N_{part}^{1/4} \approx
0.23$. This factor increases with decreasing multiplicity and,
correspondingly, centrality of the collision. We are going to give a
somewhat more careful estimate of what exactly the suppression factor
is in the framework of our model of non-flow correlations.

In principle to estimate the contribution of saturation physics to the
fourth order cumulant from \eq{c4} one has to calculate the four
particle inclusive cross section in the quasi-classical approximation
of \cite{mv,claa,kv,yuaa,km}. This task seems to be extremely
difficult. Constructing a model similar to what was done for double
inclusive cross section is dangerous since at this high order the
model inspired by factorization approaches may not be valid at
all. What we are going to do here is to neglect this four particle
correlations and estimate the contribution of two-particle
correlations of \eq{nn2} to the fourth order cumulant in \eq{c4}.

Substituting the distribution of \eq{nn2} into \eq{c4} we obtain
\ben
c_2 \{ 4 \} \, = \, 2 \int \, \frac{d N_{corr}}{d\phi_1 d\phi_3} \,
\frac{d N_{corr}}{d\phi_2 d\phi_4} \, e^{2 i (\phi_1 + \phi_2 - \phi_3
- \phi_4)} 
\een
\be\label{c41}
\times \, \left( \frac{1}{\int \frac{d N} {d \phi_1} \, \frac{d N}
{d \phi_2} \, \frac{d N} {d \phi_3} \,
\frac{d N} {d \phi_4} + 6 \int   \frac{d N}
{d \phi_1} \, \frac{d N} {d \phi_2} \, \frac{d N_{corr}}{d\phi_3
d\phi_4} + 3 \, \int \, \frac{d N_{corr}}{d\phi_1 d\phi_2} \, \frac{d
N_{corr}}{d\phi_3 d\phi_4}} - \frac{1}{\left( \int \frac{d N} {d
\phi_1} \, \frac{d N} {d \phi_2} + \int \, \frac{d N_{corr}}{d\phi_1
d\phi_2} \right)^2}\right),
\ee
where the integral sign denotes integration over all azimuthal angles
to follow. The numerator is the same in both terms in \eq{c4}. It
gives the prefactor in \eq{c41}. However the denominators are not
exactly the same in the two terms in \eq{c4}, and this is reflected by
the difference in the parentheses of \eq{c41}. The denominator of the
second term is equal to the number of pairs squared, while the
denominator of the first term is equal to the number of quadruplets of
particles.

Expanding the expression in parentheses of \eq{c41} to the lowest
non-trivial order in $d N_{corr}/ d\phi_1 d\phi_2$ yields
\be\label{c42}
c_2 \{ 4 \} \, \approx \, - 8 \, \left[ \frac{\int \frac{d N_{corr}}{d\phi_1
d\phi_3} \, e^{2 i (\phi_1 - \phi_3)}}{\int \frac{d N} {d \phi_1} \,
\frac{d N} {d \phi_3}} \right]^2 \, \frac{\int \frac{d N_{corr}}{d\phi_1
d\phi_2}}{\int \frac{d N} {d \phi_1} \,
\frac{d N} {d \phi_2}} \, = \, - 8 \, c_2\{ 2\}^2 \,  
\frac{\int \frac{d N_{corr}}{d\phi_1
d\phi_2}}{\int \frac{d N} {d \phi_1} \, 
\frac{d N} {d \phi_2}} \, = \, - 8 \, v_2\{ 2\}^4 \, 
\frac{\int \frac{d N_{corr}}{d\phi_1
d\phi_2}}{\int \frac{d N} {d \phi_1} \,
\frac{d N} {d \phi_2}}
\ee
where we follow the notation of \cite{cumul} for elliptic flow
variable $v_2\{ 2\}$ extracted from the second order cumulant $c_2\{
2\}$ (see \eq{v2corr}). Using \eq{c41} we can deduce from \eq{c42} the
following expression for the elliptic flow extracted from the fourth
order cumulant
\be\label{c43}
v_2 \{ 4 \} \, = \, v_2 \{ 2 \} \, \left[ \frac{8 \, \int \frac{d
N_{corr}}{d\phi_1 d\phi_2}}{\int \frac{d N} {d \phi_1} \,
\frac{d N} {d \phi_2}} \right]^\frac{1}{4}. 
\ee
To put a lower bound on the suppression factor in \eq{c43} we first
note that since cosine is always less or equal to one
\be\label{lb}
\frac{\int \frac{d
N_{corr}}{d\phi_1 d\phi_2}}{\int \frac{d N} {d \phi_1} \,
\frac{d N} {d \phi_2}} \, \ge \, v_2\{ 2\}^2.
\ee
According to the $\sqrt{s}=130$~GeV data reported by STAR the elliptic
flow for the most central collisions is $v_2 = 1.87 \pm 0.25\%$ which
after insertion into Eqs. (\ref{lb}) and (\ref{c43}) gives the
suppression factor
\be\label{vv1}
\frac{v_2\{ 4\}}{ v_2\{ 2\}} \, \ge \, 0.23,
\ee
where the precise agreement with the number cited before is
coincidental. Using the full expression in \eq{c41} instead of
\eq{c43} does not significantly change the result.

The suppression factor of \eq{c43} can be estimated in the framework
of our minijet model \cite{KT}. The result is
\be\label{vv2}
\frac{v_2\{ 4\}}{ v_2\{ 2\}} \, \approx \, 0.5 \div 0.75,
\ee
where the discrepancy is due to infrared cutoff sensitivity of our
model. An exact calculation in the saturation framework of
\cite{mv,claa,yuaa,km,KT1} should give a more precise estimate. Both
numbers in Eqs. (\ref{vv1}) and (\ref{vv2}) are consistent with the
STAR data presented in Fig. 13 of
\cite{starsat2}. There the ratio of the elliptic flow extracted from
the fourth and the second order cumulants is in the range of
$0.25-0.5$ for the most central events, where the discrepancy is due
to different methods of calculating the fourth order cumulant
\cite{starsat2}.

Using the same two-particle correlations model we can estimate the
ratio of the elliptic flow variable extracted from the $6$th-order
cumulant to the one extracted from the $2$nd-order cumulant to be
\be
\frac{v_2\{ 6\}}{ v_2\{ 2\}} \, = \, \left[ 108 \left( \frac{\int \frac{d
N_{corr}}{d\phi_1 d\phi_2}}{\int \frac{d N} {d \phi_1} \,
\frac{d N} {d \phi_2}} \right)^2 
\right]^\frac{1}{6} \, \approx \, 0.5 \div 0.75
\ee
where we give a numerical estimate of the ratio in our model for the
most central collisions.

Fig. 13 of \cite{starsat2} shows that the ratio of $v_2\{ 4\} / v_2\{
2\}$ increases approaching unity with decreasing centrality down to
$60\%$ centrality. This is in qualitative agreement with our
\eq{c43}, which implies that $v_2\{ 4\} / v_2\{ 2\} \sim
1/N_{part}^{1/4} \sim 1/(S_\perp^A Q_s^2)^{1/4}$ so that the ratio
increases with decreasing centrality. 

The steps shown in Eqs. (\ref{c41}), (\ref{c42}) and (\ref{c43}) can
be repeated for differential elliptic flow $v_2 (p_T)$ analysis in the
framework of the same two-particle correlation model of
\eq{nn2}. Fixing the transverse momentum of one of the particles in the 
cumulant analysis to be $p_T$ we obtain the following expression for
differential elliptic flow extracted from the fourth order cumulant
$v_2 \left\{ 4 \right\} (p_T)$ in terms of the differential elliptic
flow from the second order cumulant $v_2 \left\{2\right\} (p_T)$
\be\label{v24p}
v_2 \left\{ 4 \right\}(p_T) \, \approx \, v_2 \left\{2\right\}(p_T) \
\frac{1}{2} \, \left[ \frac{8 \, \int \frac{d
N_{corr}}{d\phi_3 d\phi_4}}{\int \frac{d N} {d \phi_3} \,
\frac{d N} {d \phi_4}} \right]^\frac{1}{4} \, \left( 1 + \frac{\int \frac{d
N_{corr}}{d^2 p_T d\phi_2} \, \frac{d N} {d \phi_3}}{\int \frac{d
N} {d^2 p_T} \, \frac{d N_{corr}}{d\phi_3 d\phi_4}}\right),
\ee
where the integral sign still implies integration over all azimuthal
angles following it. At high transverse momentum $p_T$ the ratio in
the parenthesis of \eq{v24p} goes to a constant since both the
numerator and the denominator scale in the same way with $p_T$
\cite{KT}. Therefore, since in our model \cite{KT} 
$v_2 \left\{2\right\}(p_T)$ goes to a constant at high $p_T$, the
differential elliptic flow extracted from the fourth order cumulant
$v_2 \left\{4\right\}(p_T)$ would also approach a constant at high
$p_T$ in qualitative agreement with the STAR analysis
\cite{starsat2}. To estimate the value of this $v_2 \left\{4\right\}(p_T)$ 
asymptotics we employ \eq{vv2} to write
\be\label{v24p1}
\frac{v_2 \left\{4\right\}(p_T)}{v_2 \left\{2\right\}(p_T)} \, \approx \, 
(0.5 \div 0.75) \, \frac{1 + \ln Q_s/\Lambda}{2} \, \approx \, 0.7
\div 1.0 \hspace*{1cm} (\mbox{high} \ p_T)
\ee
for $Q_s = 1$~GeV and infrared cutoff $\Lambda = 0.15$~GeV. The
numbers shown in \eq{v24p1} appear to be consistent with the STAR
results shown in Fig. 15 of \cite{starsat2}.
 
Despite the qualitative (and semi-quantitative) agreement of our model
with the data of \cite{starsat2} we do not attempt to fit the STAR
fourth order cumulant data at the moment since for a consistent fit
one needs to include the contributions of the $4$-particle
correlations in \eq{c4}.

\section*{Acknowledgments}

The authors are indebted to Kirill Filimonov, Miklos Gyulassy, Ulrich
Heinz, Dmitri Kharzeev, Roy Lacey, Larry McLerran, Jean-Yves
Ollitrault, Jan Rak, Edward G. Sarkisyan, Edward Shuryak, Andrei
Starinets, George Sterman, Derek Teaney, Raju Venugopalan, and Urs
Wiedemann for many interesting and informative discussions.

The work of Yu. K. was supported in part by the U.S. Department of
Energy under Grant No. DE-FG03-97ER41014 and by the BSF grant $\#$
9800276 with Israeli Science Foundation, founded by the Israeli
Academy of Science and Humanities. The work of K. T. was sponsored in
part by the U.S. Department of Energy under Grant
No. DE-FG03-00ER41132.

\end{document}